\UseRawInputEncoding

\documentclass[preprint,review,12pt]{elsarticle}

\usepackage{graphicx}
\usepackage{amssymb}

\usepackage{lineno}

\usepackage{soul}

\usepackage{xcolor}
\usepackage{amsmath}

\newcommand{\boldm}{\mbox{\boldmath$m$}}

\journal{International Journal of Solids and Structures}

\begin{document}
	
\begin{frontmatter}
		
\title{Phase-field modeling of paramagnetic austenite - ferromagnetic martensite transformation coupled with mechanics and micromagnetics}
		
	
		
\author[mymainaddress]{Dominik Ohmer\corref{mycorrespondingauthor}}
\cortext[mycorrespondingauthor]{Corresponding author. E-mail: dohmer@mfm.tu-darmstadt.de}
\author[mysecondaryaddress]{Min Yi\corref{mysecondcorrespondingauthor}}
\cortext[mysecondcorrespondingauthor]{E-mail: yimin@nuaa.edu.cn}
\author[mymainaddress]{Oliver Gutfleisch}
\author[mymainaddress]{Bai-Xiang Xu}
		
\address[mymainaddress]{Institute of Materials Science, Technische Universität Darmstadt, 64287 Darmstadt,
		Germany}
\address[mysecondaryaddress]{State Key Lab of Mechanics and Control of Mechanical Structures \& Key Lab for Intelligent Nano Materials and Devices of Ministry of Education \& College of Aerospace Engineering,
		Nanjing University of Aeronautics and Astronautics (NUAA), 210016 Nanjing,
		China}
		
\begin{abstract}
	A three-dimensional 
	phase-field model is proposed for simulating the magnetic martensitic phase transformation. 
	The model considers a paramagnetic cubic austenite to ferromagnetic tetragonal martensite transition, as it occurs in magnetic Heusler alloys like Ni$_2$MnGa, and is based on a Landau 2-3-4 polynomial with temperature dependent coefficients. The paramagnetic-ferromagnetic transition is recaptured by interpolating the micromagnetic energy as a function of the order parameter for the ferroelastic domains. The model is numerically implemented in real space by finite element (FE) method. FE simulations in the martensitic state show that the model is capable to correctly recapture the ferroelastic and -magnetic microstructures, as well as the influence of external stimuli. Simulation results indicate that the paramagnetic austenite to ferromagnetic martensite transition shifts towards higher temperatures when a magnetic field or compressive stress is applied. The dependence of the phase transition temperature shift on the strength of the external stimulus is uncovered as well. Simulation of the phase transition in magnetocaloric materials is of high interest for the development of energy-efficient magnetocaloric cooling devices.
\end{abstract}
		
		
\end{frontmatter}
	
	
\section{Introduction}\label{sec.Introduction}
	
With the increasing global prosperity, the energy consumption related to cooling application is increasing and expected to surpass the energy consumption related to heating within the next decades.~\cite{Isaac2009} Despite this trend, the working principle for cooling devices, which is based on gas compression, has not changed significantly for over 100 years.~\cite{Belman-Flores2015} Recent research in this field is focused on solid-state refrigeration utilizing the magneto-,~\cite{Krenke2005,Liu2012a} electro-,~\cite{Moya2014} or elastocaloric~\cite{Bonnot2007,Tusek2015} effect. The caloric effects are based on the change of the material's entropy due to a magnetic, electric, or elastic field, respectively. 
Especially since the discovery of the giant magnetocaloric effect (MCE) \cite{Pecharsky1997}, magnetocaloric cooling systems are one of the most promising alternatives to conventional gas compression cooling systems. However, up to date no commercially competitive magnetic refrigerator has been produced.~\cite{Kitanovski2015} One of the major challenges is the hysteresis in materials with large magnetocaloric effects, like Heusler alloys.~\cite{Gutfleisch2016,Taubel2018,Scheibel2018} 
Gottschall \textit{et al.} proposed a multi-stimuli concept, which is based on the application of multiple external stimuli to control the state of the material.~\cite{Gottschall2018} 
	
As the construction of a device based on the multi-stimuli concept and the synthesis of new materials with large magnetocaloric effects that are also susceptible to a mechanical stimulus could be difficult, time-consuming, and expensive, there is the need to support the research by modelling the concept. Several phase-field models exist describing ferromagnetic shape memory alloys (FMSMA). A conventional phase-field model has been developed, which uses transformation strains as order parameter.~\cite{Koyama2003,Koyama2008,Zhang2005,Wu2008,Wu2011,Wu2016,Bouville2008} It is based on an polynomial energy expansion in terms of transformation strains and magnetization, which leads to many expansion coefficients that need to be fine-tuned. The conventional phase-field model was further developed by Li \textit{et al.}~\cite{Li2008,Li2011a} expressing the transformation strain by a set of characteristic functions of the variants introduced by Shu \textit{et al.}.~\cite{Shu2007,Shu2008} In this way, the number of order parameters can be reduced and one could numerically yield multi-rank laminated domain structure. A similar approach, using free energy formalism, was used by Mennerich \textit{et al.}.~\cite{Mennerich2011,Mennerich2011} 
	
In 2016, Wu \textit{et al.}~\cite{Wu2016} and Yi \textit{et al.}~\cite{Yi2016a} published phase-field models for the microstructure of FMSMA, using real-space numerical methods. Previously, models were implemented using spectral methods like the Fast-Fourier transform (FFT), which require periodic boundary conditions that limit the system size to be infinite and depend on reciprocal-space calculation and further inverse Fourier transform for obtaining real-space values.~\cite{Jin2009,Li2008,Wu2011,Chen2002,Zhang2005,Hu2011,Mennerich2013} Wu's model was a finite element implementation of his model published earlier~\cite{Wu2011}. Yi \textit{et al.} constructed a constraint-free phase model using order parameters motivated by a multi-rank laminated microstructure for the ferroelastic ordering and azimuthal angles for the ferromagnetic orderings.~\cite{Yi2014} By this, the constraints on magnetization magnitude and the sum of volume fractions are satisfied automatically.~\cite{Yi2015,Yi2015a,Yi2015b,Yi2015c} 
The finite element implementation of the models can be used to investigate the formation and evolution of ferroelastic and ferromagnetic domains in different geometries and under different boundary conditions. The finite element implementation of the model has been extended in the recent work~\cite{Dornisch2019}.

The above reviewed models are only focused on the martensite reorientation and the influence of field and pressure on it. These models are not constructed to simulate the austenite-martensite phase transition. Models also considering the austenite-martensite phase transition (PT) are based on the work of Wang and Khachaturyan~\cite{Khachaturyan1997} who present the first three-dimensional phase-field model for generic cubic to tetragonal transitions. Levitas \textit{et al.} developed a three-dimensional Landau theory for stress- and temperature-induced austenite-martensite phase transitions.~\cite{Levitas,Levitasa,Levitasb} Since then several studies also considered hexagonal-to-orthorhombic,~\cite{Wen2000} cubic-to-tetragonal,~\cite{Jin2001} or tetragonal-to-monoclinic~\cite{Mamivand2013a} transitions. These models have been constantly modified in order to recapture the influence of stress~\cite{Artemev2001,Artemev2001} or magnetic field,~\cite{Koyama2003} on the martensite reorientation. However, most of these studies only focus on the martensitic state. Man \textit{et al.}~\cite{Man2011} studied the forward and reverse martensite phase transformation (MPT) between the martensite and austenite state with continuously varying the temperature, showing the evolution of the austenite and martensite phase up on cooling and heating. Malik and Yeddu intensively studied stress- and temperature-induced MPT in steels.~\cite{Malik2012,Malik2013,Malik2013a,Yeddu2012,Yeddu2012a,Yeddu2013,Yeddu2013b} Their work was further continued by Cui \textit{et al.}~\cite{Cui2017} who investigated the forward and reverse transformation considering the latent heat of transformation.
The presented models either focus on the influence of stress and field on martensite reorientation or on the the influence of stress and temperature on the austenite-martensite phase transition. 
	
In this work, we aim to develop a three-dimensional phase-field model for a ferromagnetic tetragonal martensite and paramagnetic cubic austenite system. Typical systems exhibiting this properties are magnetic Heusler alloys like the well-known FMSMA Ni$_2$MnGa.~\cite{Ooiwa1992,Brown1999,Ma2000} The model is capable of capturing the MPT, as well as the austenite-martensite PT, and the influence of stress and magnetic field on both. The formulation of the chemical energy is based on a Landau 2-3-4 polynomial with temperature dependent coefficients. The magnetic energy contributions are considered via micromagnetic formulations. In order to capture the ferromagnetic martensite to paramagnetic austenite PT, we scale the magnetic energy contributions with the order parameter representing the martensite/austenite state.  We first show that the proposed model is capable of replicating microstructural features in the martensitic state which are also obtained by other three-dimensional models for similar systems. We then continue to simulate the paramagnetic austenite to ferromagnetic martensite PT by performing isothermal simulations with step wise decrease of temperature. In addition, we show that the temperature at which the PT occurs can be shifted towards higher values by applying mechanical loading or magnetic field. We also show the dependence of the phase transition shift on the strength of the external stimuli.
	
\section{Theory}
\label {sec.Theory}
	
\subsection{Phase-field model}
	
For the description of martensitic transformation, order parameters describing the martensite variants and the austenite phase have to be introduced. In this work, we consider a ferromagnetic, tetragonal martensite to paramagnetic, cubic austenite transformation, which \textit{e.g.} can be found in Ni$_2$MnGa. The three martensite variants are described by the order parameters $\eta_i$, where $i$ is the number of martensite variants (three in this case). The austenite state is also described when all $\eta_i$ are zero. Using these order parameters, the chemical energy $F^\text{chem}$ of the system can be expressed as a Landau 2-3-4 polynomial~\cite{Jin2009,Malik2013,Yeddu2013,Cui2017}
	
\begin{equation}
	\begin{split}
		F^\text{chem} &= \int_V \left[ \frac{A}{2}\left(\eta_1^2+\eta_2^2+\eta_3^2\right) - \frac{B}{3}\left(\eta_1^3+\eta_2^3+\eta_3^3\right) + \frac{C}{4} \left(\eta_1^2+\eta_2^2+\eta_3^2\right)^2 \right] dV \,,
	\label{eq.F_chem}
	\end{split}
\end{equation}
where $A=32\Delta G^*$, $B = 3A-12\Delta G_\text{m}$, and $C = 2A-12\Delta G_\text{m}$. The parameters of the polynomial can be expressed by the energy barrier $\Delta G^*$ between the austenite and the martensite and the driving force $\Delta G_\text{m}$. Yeddu \textit{et al.} derived $\Delta G^*=\frac{V_\text{m}\beta}{2\delta^2}$, with $V_\text{m}$ as molar volume, $\delta$ as thickness of the interface, and $\beta$ as gradient coefficient.~\cite{Yeddu2012} The driving force $\Delta G_\text{m}$ is related to the Clausius-Clapeyron equation and its temperature dependence can be expressed as $\Delta G_\text{m}=Q(T-T_0)/T_0$, with $Q$ as the latent heat of transformation, and $T_0$ as the transition temperature.~\cite{Zhang2007} For the temperature dependency of $\Delta G^*$ we follow Cui \textit{et al.}~\cite{Cui2017} with 
	
\begin{equation}
		\Delta G^* = \quad \left\{
		\begin{aligned} &\frac{0.3}{32}Q\,, && T\leq T_0 \\ &\frac{0.8+0.06(T-T_0)}{32}Q\,, && T>T_0\,.\end{aligned}\,
		\right.
	\label{eq.deltaG}
\end{equation}
In order to constrain $\eta_i$ and $\sum_{i=1}^{3}\eta_i$ to be either 0 or 1 , we add two penalty terms to the chemical energy
	
\begin{equation}
	F^\text{pen} = \int_V P_1 \sum_{i=1}^3 \eta_i^2(1-\eta_i)^2 + P_2\left( 1-\sum_{i=1}^3\eta_i \right)^2 dV \,.
\end{equation} 
The energy at the interface of martensite variants is described by the gradient energy $F^\text{grad}$
	
\begin{equation}
	F^\text{grad} = \frac{\beta}{2} \int_V \sum_{i=1}^3\left( \nabla\eta_i \right)^2 dV \,,
\end{equation}
where $\beta$ is the gradient energy coefficient and is related to $\Delta G^*$.
The elastic energy $F^\text{ela}$ takes into account stress and strain arising from the martensitic transformation and and martensite reorientation of the variants and can be described by
	
\begin{equation}
	F^\text{ela} = \frac{1}{2}\int_V \mathbb{C}_{ijkl}\varepsilon^\text{ela}_{ij}\varepsilon^\text{ela}_{kl} dV \,, \quad \text{with} \quad \varepsilon_{ij}^\text{ela} = \varepsilon_{ij} - \varepsilon_{ij}^0 \quad \text{and} \quad \varepsilon_{ij}^0  = \sum_{s=1}^3\varepsilon_{ij}^{00}(s)\eta_s \,,
\end{equation}
where $\mathbb{C}$ is the elastic stiffness tensor. In this work, we assume an isotropic elasticity tensor which is independent of the temperature. In real systems, the elasticity tensor is a function of temperature and a change of the elastic properties from the martensite to austenite phase can occur. However, consideration of temperature-dependent elastic properties is saved for future works.
The tetragonal transformation strain is described by $\varepsilon_{ij}^{00}(s)$ with
	
\begin{equation}
	\varepsilon_{ij}^{00}(1) = 
	\begin{pmatrix}
		\varepsilon_3 & 0 & 0 \\
		0 & \varepsilon_1 & 0 \\
		0 & 0 & \varepsilon_1
	\end{pmatrix}
	\,,\quad
	\varepsilon_{ij}^{00}(2) = 
	\begin{pmatrix}
		\varepsilon_1 & 0 & 0 \\
		0 & \varepsilon_3 & 0 \\
		0 & 0 & \varepsilon_1
	\end{pmatrix}
	\,,\quad
	\varepsilon_{ij}^{00}(3) = 
	\begin{pmatrix}
		\varepsilon_1 & 0 & 0 \\
		0 & \varepsilon_1 & 0 \\
		0 & 0 & \varepsilon_3
	\end{pmatrix}\,.
\end{equation}

It should be noted that $P_1$ and $P_2$ will affect the effective elastic tensor, as demonstrated by Landis.~\cite{Landis2008} The difference between the measured elastic tensor and the elastic stiffness given in Eq.\,5 and Eq.\,12 depends on the curvature of the energy wells. In this work, the curvature is assumed to be sufficiently large to assume that this difference should have a negligible influence on the microstructure evolution.  The introduction of elastic tensor as a function of $P_1$ and $P_2$ should be explored in the near future.
	
\subsection{Micromagnetism}\label{sec.Micromagnetism}
For the description of the magnetic energy of the system we employ micromagnetic formulations. The magnetic energy $F^\text{mm}$ is given as
	
\begin{equation}
	F^\text{mm} = F^\text{exc} + F^\text{ani} + F^\text{mag} + F^\text{ms}\,,
	\label{eq.magneticEnergy}
\end{equation}
where $F^\text{exc}$ is the exchange energy, $F^\text{ani}$ the anisotropy energy, $F^\text{mag}$ the magnetostatic energy, and $F^\text{ms}$ the elastic energy due to magnetostriction.
Micromagnetic simulations assume constant saturation magnetization $M_\text{s}$, which is not the case considering the temperature range we are interested in, especially for a transition from magnetic martensite phase to paramagnetic austenite phase. In order to consider the change of  $M_\text{s}$ and thus the change of the magnetic energies, we scale the magnetization related energy terms with $\left(\eta_1^2+\eta_2^2+\eta_3^2\right)$. This term is either 0 or 1, where 1 represents the magnetic martensite state and 0 the paramagnetic austenite state. Based on this approach, the exchange energy $F^\text{exc}$ is given as
	
\begin{equation}
	F^\text{exc} = \int_V \left(\eta_1^2+\eta_2^2+\eta_3^2\right) A_\text{e}||\nabla\mathbf{m}||^2 dV\,, 
\end{equation}
with $A_\text{e}$ as exchange stiffness related to the Curie temperature $T_\text{C}$ and $\mathbf{m}$ as the magnetization unit vector. 
The anisotropy energy $F^\text{ani}$ penalizes the misalignment of magnetization with the magnetic easy axis $\boldsymbol{e}^\text{a}$ of the system. $F^\text{ani}$ is zero for parallel alignment with the easy axis and at maximum if magnetization and easy axis are perpendicular. In FMSMA, the easy axis depends on the martensite variant, \textit{e.g.}, in Ni$_2$MnGa the easy axis is parallel to the short axis of the tetragonal martensite variant. Thus, for the cubic to tetragonal transition considered here, the easy axis as a function of the order parameters $\eta_i$ can be given as
	
\begin{equation}
	\boldsymbol{e}^a(\eta_i) = \sum_{i=1}^3\eta_i \boldsymbol{e_i}\ = \eta_1\begin{pmatrix}1\\0\\0\end{pmatrix} + \eta_2\begin{pmatrix}0\\1\\0\end{pmatrix} + \eta_3\begin{pmatrix}0\\0\\1\end{pmatrix}\,.
\end{equation}
From this, $F^\text{ani}$ can be expressed by
\begin{equation}
	F^\text{ani} = \int_V \left(\eta_1^2+\eta_2^2+\eta_3^2\right) K_1 \left( 1- \|\boldsymbol{m}\cdot\boldsymbol{e}^\text{a}(\eta_i)\|^2 \right) dV \,,
	\label{eq.coupling}
\end{equation}
with $K_1$ as the anisotropy constant. 
The anisotropy energy is the main coupling term between the ferroelastic and ferromagnetic domain. This coupling mechanism enables the system to change the magnetization by applying pressure or to transform the martensite variants by applying an external magnetic field.
The magnetostatic energy term $F^\text{mag}$ takes into account demagnetization field and Zeeman energy
	
\begin{equation}
	F^\text{mag} =  \int_V \left[\frac{1}{2}\mu_0\mathbf{H}_d\cdot\mathbf{H}_d-\mu_0\left(\eta_1^2+\eta_2^2+\eta_3^2\right)M_\text{s}\mathbf{H}\cdot\mathbf{m}\right]dV\,,
\end{equation}
where $\boldsymbol{H}_d$ is the demagnetization field and $M_\text{S}$ is the saturation magnetization. Although the contribution of the elastic energy due to magnetostriction may be small compared to the other energy terms, it is considered here for the completeness of the model, which leads to an overall elastic energy described by
	
\begin{equation}
	F^\text{ms}= \int_V \frac{1}{2}\mathbb{C}_{ijkl}\left(\varepsilon_{ij}-\varepsilon^0_{ij}-(\eta_1^2+\eta_2^2+\eta_3^2)\varepsilon_{ij}^\text{ms}\right)\left(\varepsilon_{kl}-\varepsilon^0_{kl}-(\eta_1^2+\eta_2^2+\eta_3^2)\varepsilon_{kl}^\text{ms}\right) dV\,,
\end{equation}
with $\boldsymbol{\varepsilon}^\text{ms}$ as the magnetostrictive strain tensor. According to the assumption of isotropic elasticity, we further assume isotropic magnetostriction with
	
\begin{equation}
	\varepsilon^\text{ms}_{ij}=\frac{3}{2}\lambda_s\left(m_im_j -\frac{1}{3}\delta_{ij}\right) \,,
\end{equation}
where $\lambda_s$ is the magnetostrictive constant.
	
\subsection{Governing equations}
The time evolution of the ferroelastic domains, described by the order parameters $\eta_i$, is modeled using the Allen-Cahn equation,~\cite{Allen1979} also known as time-dependent Ginzburg-Landau (TDGL) equation~\cite{khachaturyan2013theory}
	
\begin{equation}
	\frac{\partial \eta_p}{\partial t} = - L \frac{\partial F}{\partial\eta_p}\,,
	\label{eq.TDGL}
\end{equation}
with $L$ as the kinetic parameter and $F$ as the free energy of the system, which is the sum of all energy terms introduced above. The evolution of magnetization over time is given by the Landau-Lifshitz-Gilbert (LLG) equation
	
\begin{equation}
	\dot{\boldsymbol{m}} = -\mu_0 |\gamma| \boldsymbol{m} \times \boldsymbol{H}_\text{eff} + \alpha \boldsymbol{m} \times \dot{\boldsymbol{m}} \,,
	\label{eq.LLG} 
\end{equation}
where $|\gamma|$  is the gyromagnetic ratio and $\alpha$ is the Gilbert damping constant.
The effective magnetic field $\boldsymbol{H}_\text{eff}$ is given as
\begin{equation}
	\boldsymbol{H}_\text{eff} = -\frac{1}{\mu_0 M_\text{s}}\frac{\delta F}{\delta\mathbf{m}}\,.
\end{equation}
For the micromagnetic formulations, the magnitude of the magnetization is given by the saturation magnetization $M_\text{s}$. Therefore, the magnetization vector $\boldsymbol{m}$ has to follow the constraint
	
\begin{equation}
	\|\boldsymbol{m}\|^2-1=0\,.
\end{equation}
This constraint is implemented using the Lagrangian multiplier method
	
\begin{equation}
	\Pi_c = \int_V \lambda (\|\boldsymbol{m}\|^2-1) d\Omega - \int_\Omega \frac{\lambda^2}{2k} dV\,,
\end{equation}
where $\lambda$ is the Lagrange multiplier and k is a constant on the order of $10^{5}$.

With the model including an elastic energy contribution that couples the magnetization and order parameters $\eta_i$ with displacements, the mechanical equilibrium given by
\begin{equation}
	\nabla \cdot \mathbf{\sigma} = \sigma_{ij,i} = 0
\end{equation}
has to be considered as well. In addition, including magnetism further requires considering the magnetic equilibrium given by the Maxwell relation
\begin{equation}
	\nabla \cdot \mathbf{B} = \mathbf{B}_{i,i} = 0 \,,
\end{equation}
with
\begin{equation}
	\mathbf{B} = \mu_0\left(\mathbf{H}+M_\text{s}\mathbf{m}\right) \,.
\end{equation}
The magnetic field $\mathbf{H}$ is defined via the scalar magnetic potential $\phi$
\begin{equation}
	\mathbf{H} = -\nabla \phi \,,
\end{equation}
which is an additional degree of freedom that can be used to calculate the demagnetization field and magentostatic energy.

\section{Calculation details}\label{sec.CalculationDetails}
	
For the numerical implementation of the model, we write our own finite-element code based on the open-source Multiphysics Object-Oriented Simulation Environment (MOOSE).~\cite{Permann2020} In the MOOSE framework, coupling between different physics is straightforward.~\cite{Tonks2012,Novascone2015,Tonks2016} MOOSE is build on the meshing and finite element library of libmesh~\cite{Kirk2006} and the non-linear solver PETSc~\cite{abhyankar2018petscts}. For the validation of the proposed model, we use model parameters which are in the range of typical Heusler alloys of interest, like Ni$_2$MnGa, and can be found in Table\,\ref{tab.parameters}. The transition temperature $T_0$ is set as 200\,K. As can be seen from $E$ and $\nu$, we assume an isotropic elasticity stiffness tensor $\mathbb{C}$ for the martensite and austenite phase.
\begin{table}[!htb]
	\centering
	\caption{Model parameters used within this work.}
	\begin{tabular}{c|c|c|c|c}
		$E$ [GPa] & $\nu$ [GPa] & $\varepsilon_1$ & $\varepsilon_3$ & $\lambda_s$\\\hline
		$150$     & $0.3$       & $0.02$          & $-0.03$ & $150\cdot 10^{-6}$\\
		& & & \\
		$M_\text{s}$ [A/m]& $K$ [J/m$^3$]   & $A_e$ [J/m]       & $\beta$ [J/m]        & $Q$ [J/m$^3$] \\\hline
		$5\cdot 10^5$     & $2.5\cdot 10^5$ & $2\cdot 10^{-11}$ & $1.5\cdot 10^{-10}$  & $7\cdot 10^5$ \\
	\end{tabular}
	\label{tab.parameters}
\end{table}
	
\section{Results}\label{sec.Results}
In the following, we show that the proposed model is capable of capturing the martensitic state with multi-variant microstructures, together with the dependence of the structure on boundary conditions and geometry. We show the ferroelastic and ferromagnetic coupling in a single-element benchmark and a microstructure. The austenite-martensite transition is simulated by the step-wise reducing the temperature. Repeating this for applied magnetic field and pressure shows a shift in the transition temperature towards higher temperatures.
	
\subsection{Single-element benchmark}
	
\begin{figure}[!htb]
	\centering
	\includegraphics[width=\textwidth]{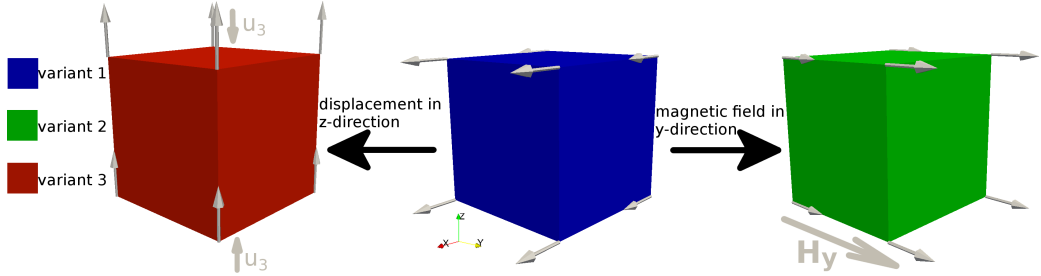}
	\caption{Single-element benchmark showing that the martensite variant can be transformed by applying stress/strain or an external magnetic field.}
	\label{fig.SE_benchmark}
\end{figure}

We perform a single-element benchmark to show the coupling between the order parameters $\eta_i$ and the magnetization unit vector $\boldm$. The negative sign of $\varepsilon_3$ indicates that the tetragonal martensite exhibits one shorter axis, which in this model are oriented in direction of the unit axes. For $\eta_1$ the short axis is the $x$-axis, for $\eta_2$ it is the $y$-axis and for $\eta_3$ the $z$-axis. The short axis of the martensite variant is also the magnetic easy axis of the corresponding variant. For the single-element benchmark the initial martensite variant is chosen to be $\eta_1$, see middle of Fig.\,\ref{fig.SE_benchmark}. By applying a strain, equal to $\varepsilon_3$ or larger, in $z$-direction the martensite transforms from $\eta_1$ to $\eta_2$. The magnetization follows and rotates into the $z$-axis direction. The strain is exerted on the element by applying a displacement in $z$-direction while the other surfaces are not constrained (free to move). Similar behavior is obtained if a sufficient magnetic field is applied. In analogy to the strain application, we apply a magnetic field in $y$-direction. The magnetization rotates towards the $y$-direction and then transforms from $\eta_1$ to $\eta_2$ which is accompanied by shrinkage in $y$-direction. For the application of the magnetic field, all surfaces are free to move. 
Despite similar behavior, it is important to note the different mechanisms for the stress- and field-induced martensite reorientation. For the application of strain, the elastic energy of the system increases. The system transforms from $\eta_1$ to $\eta_3$ in order to reduce the increased elastic energy again. The rotation of the magnetization then follows in order to further reduce the overall energy of the system by reducing the magnetocrystalline anisotropy energy. For the application of an external magnetic field, the order is reversed. With the applied field the magnetocrystalline anisotropy energy increases. By transforming from $\eta_1$ to $\eta_2$ the system reduces the anisotropy energy. This reorientation is accompanied by the structural change from one tetragonal variant to another. Thus, the system expands in $x$-direction and compresses in $y$-direction. It is at this point mentioned that the anisotropy constant $K_1$ was significantly increased in this benchmark. The $K_1$ value given in Tab.\,\ref{tab.parameters} is too small to induce a martensite reorientation. The main purpose of this single element benchmark is to show the principle coupling between order parameters via the anisotropy energy.
	
\subsection{Influence of boundary conditions and geometry on microstructure}
\begin{figure}[!htb]
	\centering
	\includegraphics[width=.7\textwidth]{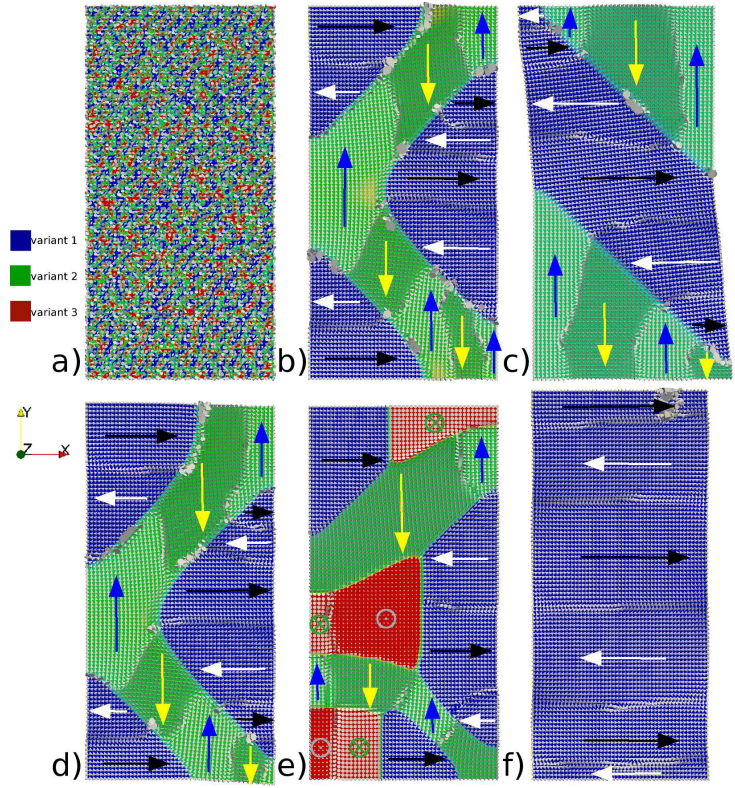}
	\caption{(a) Initial and equilibrium microstructure for a 240nm$\times$480nm$\times$5nm domain with (b) all surfaces fixed, (c) surfaces perpendicular to $x$ free, (d) surface perpendicular to $y$ free, (e) surface perpendicular to $z$ free, and (f) all surfaces free. The arrows indicate the in-plane orientation of magnetization, while the circles with a cross indicate the out-of-plane orientation.}
	\label{fig.BCs}
\end{figure}

The proposed model is applied to a 240\,nm$\times$480\,nm$\times$5\,nm domain with different boundary conditions. 
Using the model parameters shown in Table\,\ref{tab.parameters}, the magnetic exchange length $\sqrt{2A_\text{e}/(\mu_0M_\text{s}^2)}$ is around 11\,nm and the domain wall width $\pi\sqrt{A_\text{e}/K}$ is around 28\,nm. We therefore choose the mesh size to be 5$\times$5$\times$5\,nm. Starting from random distribution of $\eta_i$ and random magnetization orientation, we let the system converge to the equilibrium state for fixed and free boundary conditions, \textit{i.e.} allowing the surfaces to move or not to move, respectively (Fig.\,\ref{fig.BCs}b and f). In addition, we let the system converge allowing the surfaces perpendicular to the unit vectors to move (Fig.\,\ref{fig.BCs}c, d, and e). Fig.\,\ref{fig.BCs}b shows the equilibrium microstructure for all surfaces being fixed. The system transforms to a multi-variant structure of $\eta_1$ and $\eta_2$ with the interfaces being oriented in a 45$^{\circ}$ angle, reducing the elastic energy. $\eta_3$ disappears due to the thickness of the domain of only 5\,nm. For the fully clamped system only 90$^{\circ}$ magnetic domain walls are observed at the interfaces of $\eta_1$ and $\eta_2$. It is mentioned that in this work only the demagnetization field within the magnetic material is considered to reduce computation time. With the scalar potential $\phi$ as additional degree of freedom, the demagnetization field and magnetostatic energy can be obtained by extension of the integration domain to include the outside of the material. However, as the focus of this work is on the austenite-martensite transition the demagnetization field outside the material is neglected. We are aware that doing so can affect the resulting variant distribution and that the integration domain should be extended in future works.

When the surfaces perpendicular to $x$ are free to move (Fig.\,\ref{fig.BCs}c), the system not only reduces the elastic energy, it also reduces the interface energy ($F^\text{grad}$) by forming a larger $\eta_1$ domain between two $\eta_2$ domains. 90$^{\circ}$ domain walls are again found at the interfaces of $\eta_1$ and $\eta_2$. Within the domains of $\eta_1$ and $\eta_2$ 180$^{\circ}$ domain walls are observed. A similar structure is found for the surfaces perpendicular to $y$ being free to move (Fig.\,\ref{fig.BCs}d). When the surfaces perpendicular to $z$ are free, the system evolves to a multi-variant structure including $\eta_3$ (Fig.\,\ref{fig.BCs}e). In analogy to Fig.\,\ref{fig.BCs}c and d, 90$^{\circ}$ domain walls are observed at the interfaces of martensite variants and 180$^{\circ}$ domain walls are observed within the variants. If all surface are free to move (Fig.\,\ref{fig.BCs}f), the system forms a single variant with uniform magnetization, minimizing the interface and magnetic exchange energy. The relations between boundary conditions and resulting microstructure agree well with the results of Yi~\cite{Yi2016a}. The observation of 90$^{\circ}$ and 180$^{\circ}$ domain walls between and within variants, respectively, agrees with the work of Wu~\cite{Wu2016} as well. From these results it can already be seen that the elastic energy is the main driving force for the multi-variant structures in (FM)SMA.

\begin{figure}[!htb]
	\centering
	\includegraphics[width=\textwidth]{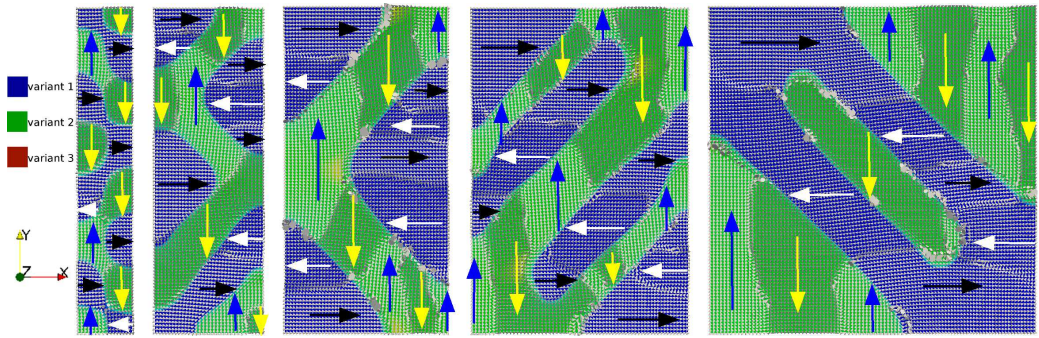}
	\caption{Equilibrium microstructure for domains with aspect ratio (a) 6, (b) 3, (c) 2, (d) 1.5, and (e) 1. Simulations were started from random distribution of $\eta_i$ and random orientation of $\boldm$ with all surfaces being fixed.}
	\label{fig.geometries}
\end{figure}

In contrast to spectral methods, using real-space numerical methods exempts from being restricted to periodic boundary conditions. This enables investigating the influence of sample geometry on the microstructure. In analogy to Yi~\cite{Yi2016a}, simulations with aspect ratios (ratio of height to width) of 6, 3, 2, 1.5, and 1 are performed (Fig.\,\ref{fig.geometries}). All simulations are started from random distribution of $\eta_i$ and random orientation of $\boldm$. For this part, all surfaces were fixed. It can be seen that the resulting microstructure strongly depends on the geometry. For a larger aspect ratios, the system forms small variant domains which alternate along $x$- and $y$-direction, minimizing the elastic energy. Minimization of the elastic energy occurs at the cost of the interface energy which is increased due to the smaller domains, increasing the interface-to-volume ratio. For decreased aspect ratios, the system is able to form larger domains of the variants, decreasing the interface energy again. The simulation of the microstructure with respect to the aspect ratio shows the competition between the elastic and interface energy.
	
\subsection{Stress-induced martensite reorientation}
\begin{figure}[!htb]
	\centering
	\includegraphics[width=\textwidth]{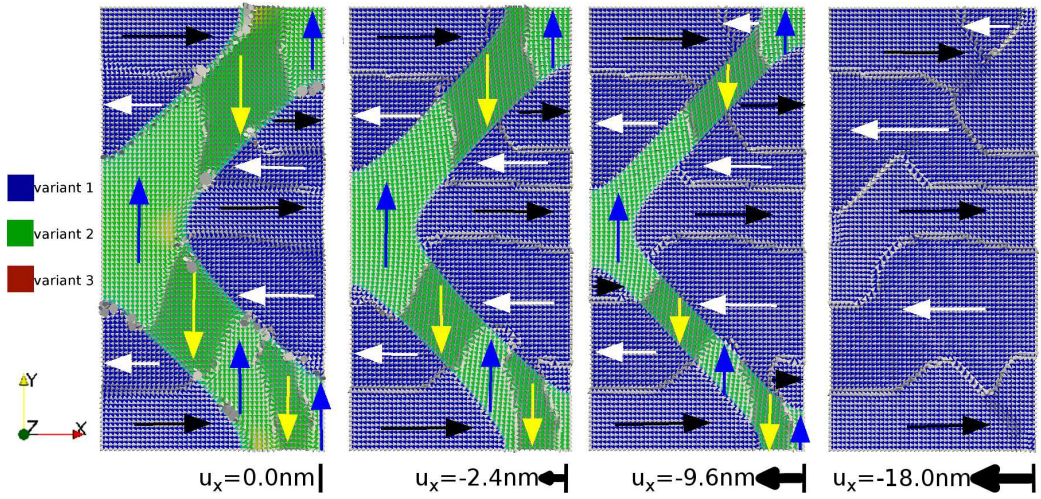}
	\caption{Microstructure evolution for compression along the $x$-direction. Structures are shown for $u_x=0, -2.4, -9.6, -18\,nm$.}
	\label{fig.SIMT_x-direction}
\end{figure}

To show the response of the system to mechanical loading in the martensite state (low temperatures), we simulate the evolution of the microstructure under compressive conditions. Using the equilibrium microstructure of the 240\,nm$\times$480\,nm$\times$5\,nm domain with clamped surfaces (Fig.\,\ref{fig.BCs}b), compression in $-x$ direction is applied. For application of the compression, displacement boundary conditions are applied on the right surface. Fig.\,\ref{fig.SIMT_x-direction} shows the evolution of martensite variants and magnetization. With increasing displacement $u_x$, the $\eta_2$ domain shrinks by getting thinner. $\eta_2$ transforms to $\eta_1$ as the crystallograpic shorter axis in $\eta_1$ is oriented parallel to the $x$-axis. By this, the system can reduce the elastic energy induced by compression. When the displacement is high enough (Fig.\,\ref{fig.SIMT_x-direction}d), $\eta_2$ is fully transformed to $\eta_1$, leading to a single $\eta_1$ domain. With the transformation, the magnetization also rotates. The magnetocrystalline easy axis changes from the $y$-axis to the $x$-axis when $\eta_2$ transforms to $\eta_1$, resulting in a rotation of the magnetization parallel to the $x$-axis. These results once again show the coupling of $\eta_i$ and $\boldm$ via the anisotropy energy, as well as the sensitivity of the system to an external stimuli.
	
\subsection{Stress- and field-induced shift of austenite-martensite transition temperature}
As the magnetocaloric effect is based on the entropy change of the system, it is largest in magnetostructural transitions, \textit{e.g.} found in magnetic Heusler alloys. The multi-stimuli concept~\cite{Gottschall2018} is based on controlling the austenite-martensite phase transition with external stimuli. Thus, in order to simulate the multi-stimuli concept, a model is needed that is capable of capturing the austenite-martensite transition with applied stress and magnetic field. While the multi-stimuli concept is applied to systems with an inverse MCE, the proposed model describes systems with a conventional MCE. Modification of the model for systems with an inverse MCE is planned for future works.
The proposed model is used to simulate the cooling curve for a single-element. Simulation of the cooling curve is performed by isothermal simulation at each temperature step, using the output of the previous step as input. The temperature of each temperature step is set via the coefficients $A$, $B$, and $C$ of Eq.\,\ref{eq.F_chem} which change with temperature due to the temperature-dependent $\Delta G^*$ and $\Delta G_m$.

\begin{figure}[!htb]
	\centering
	\includegraphics[width=\textwidth]{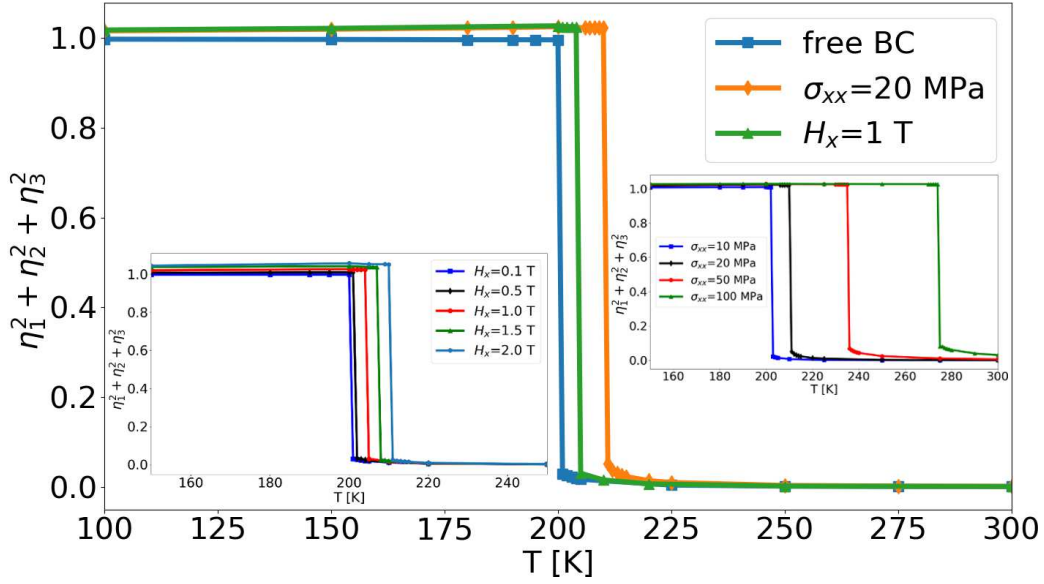}
	\caption{$(\eta_1^2+\eta_2^2+\eta_3^2)$ as a function of temperature for a single-element with free boundaries, with field applied parallel to $x$ and with pressure applied in $x$-direction. The transition temperature shifts with the application of stress or field.}
	\label{fig.PhaseTransition}
\end{figure}

For the first temperature step at 400\,K, we start with random distribution of $\eta_i$ and random orientation of $\boldm$. In order to show the austenite-martensite transition, we plot $(\eta_1^2+\eta_2^2+\eta_3^2)$ (which is 0 for the martensite and 1 for the austenite) with respect to temperature in Fig.\,\ref{fig.PhaseTransition}. The same procedure is repeated with applied stress and applied magnetic field. The results for compressive stress of 20\,MPa and for a magnetic field of 1\,T are compared to the stress- and field-free scenario in Fig.\,\ref{fig.PhaseTransition}. It can be seen that for both stimuli, the phase transition temperature shifts from 200\,K to 204\,K and 210\,K for the magnetic field of 1\,T and the compressive stress of 20\,MPa, respectively. Both external stimuli stabilize the magnetic martensite phase. The insets show the shift of the transition temperature for various magnetic fields and compressive stresses, showing linear dependence. The shift of transition temperature for applied stresses towards higher temperatures has been discovered experimentally for Ni-Mn-based Heusler alloys. For example, in the work by Gutfleisch \textit{et al.}~\cite{Gutfleisch2016}, stress-induced shifts of the transition temperature between 0.2 and 0.44\,K/MPa are shown. In our simulations, we obtain a shift of 0.5\,K/MPa, which is in the same range as the experimentally obtained values for Ni-Mn-based Heusler alloys.
It can also be seen that the system is more sensitive to mechanical loading, as the temperature shift with applied stress is significantly larger than for the magnetic fields, which are in the range of typical applications. Both the shifts with stress and field depend on the coefficients of the chemical energy, as well as on the elastic and magnetic properties of the material. For elastically harder materials (higher $E$) or a chemical energy formulation with higher energy barriers, the shift with applied stress would be decreased. In analogy, increased saturation magnetization and/or magnetocrystalline anisotropy would increase the energy of the magnetic system and thus increase the shift with respect to the applied magnetic field. 
	
\section{Conclusions}\label{sec.Conclusions}
We proposed a three-dimensional phase-field model capable of simulating the paramangetic austenite to ferromagnetic martensite phase transition. The model is based on a Landau 2-3-4 polynomial with temperature dependent coefficients for the description of the chemical energy, combined with modified micromagnetic formulations for the description of magnetism. In order to account for a magnetic to paramagnetic transition, the magnetic energy terms were modified by $(\eta_1^2+\eta_2^2+\eta_3^2)$. For the magnetic martensite $(\eta_1^2+\eta_2^2+\eta_3^2)$ is $\approx$ 1 and therefore the magnetic energy terms contribute to the overall energy of the system. In the paramagnetic austenite state $(\eta_1^2+\eta_2^2+\eta_3^2)$ is $\approx$ 0 and the magnetic energy contributions become negligible for the overall energy of the system. The proposed model is implemented in a finite-element framework and tested on a single-element benchmark, showing the coupling of the structural and magnetic properties and the ability of stress- and field-induced martensite reorientation. Simulations of multi-element structures in the martensitic state showed the dependence of ferroelastic and ferromagnetic domains on boundary conditions and sample geometry, also highlighting the FEM implementation.  We then proceeded with the simulation of a single-element for the temperature range from 400 to 0\,K and showing the phase transition at 200\,K. Simulations with applied magnetic fields and compressive stresses were performed to show the shift of the transition temperature towards higher temperatures. As we considered a magnetic martensite to paramagnetic austenite transition, both external stimuli stabilize the magnetic martensite and therefore shift the transition temperature to higher temperatures. However, materials with an inverse magnetocaloric effect, \textit{e.g.} Ni$_2$MnGa or Ni‐Mn‐X(‐Co) alloys, have a magnetic martensite to magnetic austenite phase transition, where $M_\text{s}$ of the high temperature austenite is larger than $M_\text{s}$ of the low-temperature martensite.~\cite{Taubel2018,Taubel2020} In those materials the transition temperature shifts to lower temperatures when a magnetic field is applied.~\cite{Basso2012,Gottschall2020,Pfeuffer2020,Gracia-Condal2020} In the work presented, we modified the magnetic energy terms in order to account for the magnetic martensite to paramagnetic austenite transition. By similar modification of the magnetic energy terms, the model can be extended to account for paramagnetic martensite to ferromagnetic austenite transitions (inverse MCE). It is also possible to adjust our model for the transition of a ferromagnetic martensite to a ferromagnetic austenite. For a more accurate description of the magnetic properties, temperature related magnetization can be used as input. The three-dimensional, real-space implementation of the model further enables systematic studies of the relation between the microstructure and the shift of the transition temperature, as well as the investigation of the thermal hysteresis behavior under external stimuli. By this, the multi-stimuli concept could be simulated, supporting the development of a more energy-efficient magnetocaloric cooling device. Simulation of the multi-stimuli cooling cycle can help to estimate the pressures and magnetic fields needed for the cycle, as well as creating a material property profile for magnetocaloric materials that are most suitable for the multi-stimuli cooling cylce.
	
\section{Acknowledgement}\label{sec.Acknowledgement}
The authors acknowledge the support from the European Research Council (ERC) under the European Unions Horizon 2020 research and innovation programme (Grant agreement No 743116), the NSFC (Grant No. 11902150 ),  the German Research Foundation (DFG YI 165/1-1 and DFG XU 121/7-1),  the access to the Lichtenberg High Performance Computer of Technische Universität Darmstadt, the 15th Thousand Youth Talents Program of China, the Research Fund of State Key Laboratory of Mechanics and Control of Mechanical Structures (MCMS-I-0419G01), and A Project Funded by the Priority Academic Program Development of Jiangsu Higher Education Institutions.
	
	
	



\bibliographystyle{elsarticle-num-names}
\bibliography{MicromagneticPFM_DominikOhmer_accepted_manuscript.bib}
	
\end{document}